\documentclass[preprint,showpacs,preprintnumbers,amsmath,amssymb]{revtex4}
\usepackage{graphics,epsfig,subfigure}
\usepackage{color}

\begin{document}
\renewcommand{\baselinestretch}{1.3}
\newcommand\beq{\begin{equation}}
\newcommand\eeq{\end{equation}}
\newcommand\beqn{\begin{eqnarray}}
\newcommand\eeqn{\end{eqnarray}}
\newcommand\nn{\nonumber}
\newcommand\fc{\frac}
\newcommand\lt{\left}
\newcommand\rt{\right}
\newcommand\pt{\partial}

\allowdisplaybreaks

\title{Constructing black holes in Einstein-Maxwell-scalar theory}
\author{Shuang Yu\footnote{yushuang@nao.cas.cn}\ \ \ \ Jianhui Qiu\footnote{jhqiu@nao.cas.cn}\ \ \ \ Changjun Gao\footnote{gaocj@nao.cas.cn}}

\affiliation{ Key Laboratory of Computational Astrophysics, National Astronomical Observatories, Chinese Academy of Sciences, Beijing {100101}, China}

\affiliation{School of Astronomy and Space Sciences, University of Chinese Academy of Sciences,
No. 19A, Yuquan Road, Beijing 100049, China}

\begin{abstract}

Exact black hole solutions in the Einstein-Maxwell-scalar theory are constructed. They are the extensions of dilaton black holes in de Sitter or anti de Sitter universe. As a result, except for a scalar potential, a coupling function between the scalar field and the Maxwell invariant is present. Then the corresponding Smarr formula and the first law of thermodynamics are investigated.

\end{abstract}

% \Keywords{ }



%11.10.Kk Field theories in dimensions other than four (see also 04.50.-h Higher-dimensional gravity and other theories of gravity; 04.60.Kz Lower dimensional models; minisuperspace models in general relativity and gravitation)

%04.50.Kd 	Modified theories of gravity

%04.50.-h Higher-dimensional gravity and other theories of gravity
%         (see also 11.25.Mj Compactification and four-dimensional models, 11.25.Uv D branes)

% 04.50.+h Gravity in more than four dimensions, Kaluza-Klein theory,
           % unified field theories, alternative theories of gravity
           %(see also 11.25.M Compactification and four-dimensional models), dilaton gravity

% 04.50.Gh Higher-dimensional black holes, black strings, and related objects.

% 04.70.Dy Quantum aspects of black holes, evaporation, thermodynamics.

% 11.27.+d Extended classical solutions; cosmic strings,
           %domain walls, texture (see 98.80.C in cosmology)
%98.80.-k Cosmology
%\keywords{}

\maketitle


%%%%%%%%%%%%%%%%%%%%%%%%%%%%%%%%%%%%%%%%%%%%%%%%%%%%%%%%%%%%%%%%%%%%%%%%%%%%%%%%%%%%%%%

\section{Introduction}

The theory containing coupling between the scalar fields and gravity was firstly considered by Fisher who found a static and spherically symmetric solution of the Einstein massless scalar field equations \cite{Fisher1948}. Whereafter, variety of theories {emerge} with the introduction of scalar fields, such as dimensional reductions {and} low-energy limit of string theories, various models of supergravity, etc. The low energy limit of the string theory has the Einstein action which is supplemented by fields, such as axion, gauge fields, and dilaton coupled in a nontrivial way to other fields. One of the most important theories of the low-energy limit of string theories is the {dilaton} gravitation theory \cite{green:1987}. It is a very interesting theory because the dilaton field can naturally couple to gauge fields, and the  properties of the black hole change drastically due to the dilaton field. The presence of the dilaton field can change the causal structure and asymptotic behavior of the spacetime as well as the thermodynamics and stability of the electrically charged black holes \cite{pro1}. Therefore the dilaton black holes have attracted many attention in recent decades.

On the other hand, anti-de Sitter space({AdS}) and conformal field theory correspondence (AdS/CFT) \cite{ADSCFT1} is a significant method to unify quantum fields and gravitations. This correspondence between the gravity in an AdS spacetime and conformal field theory (CFT) {leaking} on to  the boundary of the AdS spacetime suggests that physics of the gauge field defined on the AdS-boundary can be derived from gravitations in the bulk space and vice versa \cite{ADSCFT2}. Stimulated by AdS/CFT, many works are devoted to find the asymptotically (anti)-de Sitter black hole solution in dilaton gravity with various self-interacting potential of dilaton. Exact solutions of charged dilaton black holes have been constructed by many authors. Some black holes are asymptotically flat \cite{gib:1988,gar:1991,flat} where the dilaton changes the causal structure of the black hole and leads to the curvature singularities at finite radii. Meanwhile, some solutions are asymptotically neither flat nor (anti)-de Sitter \cite{nor}. Finally, the asymptotically (anti)-de Sitter versions are found in \cite{gao:2004,gao:2005a,Hajkhalili:2019,Hendi:2010,Hajkhalili:2018,Sheykhi2008,Yamazaki2001,anti}. These solutions reveal that the cosmological constant proposed in the Einstein theory is coupled to Liouville-type dilaton scalar potential.
With the building of these asymptotically (anti)-de Sitter dilaton black holes, many interesting physical phenomena have been explored, such as the black hole thermodynamics \cite{Sheykhi:2016,Sheykhi:2010,Hendi:2010}, holographic thermalization \cite{Zhang2015}, black hole phase transition in {extended} phase space \cite{Li2018} etc.

The purpose of this paper is, starting from the known dilaton black holes in de Sitter or anti-de Sitter universe, to construct new exact asymptotically (anti)-de Sitter black hole solutions in the Einstein-Maxwell-scalar theory which is the generalization of Einstein-Maxwell-dilaton theory. It is known that the Einstein-Maxwell-scalar theory can emerge naturally in physics, for example, in the contexts of Kaluza-Klein models
\cite{Ka1921}, supergravity/string theory \cite{van1981} and cosmology \cite{mar2008}. In the last two years, a remarkable phenomenon of spontaneous scalarisation of charged black holes is discovered by \cite{her2018,fer2019}. It has motivated vast studies on various Einstein-Maxwell-scalar models (see \cite{bla2020} and references therein). This is one of the motivations for this research. On the other hand, according to the no-hair theorem, there are at most three physical parameters (mass, charge and angular momentum) for a black hole. However, by {giving up} the corresponding assumptions, one can violate the no-hair theorem in the Einstein-Maxwell-scalar theory (EMS). For example, dilaton charge $D=-\frac{Q^{2}}{2M}$ emerges in diaton gravitation theory. One finds that the spacetime of dilaton black hole and {its de Sitter counterpart} are all affected by the dilaton charge, but not the electrical charge. Therefore, dilaton charge can be viewed as a new hair. In view of these points and inspired by the expression of dilaton black holes in de Sitter or anti-de Sitter universe, we find two new exact black hole solutions in the Einstein-Maxwell-scalar theory. These solutions are the extensions of Reissner-Nordstrom-(anti) de Sitter and dilaton-(anti) de Sitter solutions. We find that the new black hole spacetime in (anti)-de Sitter universe {is} directly affected by the electrical charge. {This} is different from the dilaton black hole solution.

The paper is organized as follows. In Sec. II, we derive the equations of motion in the Einstein-Maxwell-scalar theory. In Sec. III, we present the first electrically charged dilaton black hole in (anti)-de Sitter spacetime and give the analysis on its horizons. In Sec. IV, we study its thermodynamics which covers the construction of Smarr formula and first law of thermodynamics. Then the second electrically charged dilaton black hole in (anti)-de Sitter spacetime with arbitrary coupling constant $\alpha$ is found in Sec. V and  the corresponding analysis of thermodynamics are given in Sec. VI. Finally, we give the conclusion and discussion in Sec. VII. Throughout the paper, we adopt the system of units in which $G=c=\hbar=1$ and the metric signature
$(-,\ +,\ +,\ +)$.

\section{The equations of motion}
The action of Einstein-Maxwell-scalar theory is given by
\begin{equation}
S=\int d^4x\sqrt{-g}\left[R-2\nabla_{\mu}\phi\nabla^{\mu}\phi-K\left(\phi\right)F^2-V\left(\phi\right)\right]\;,
\end{equation}
where $R$ is the Ricci scalar curvature, $F^2\equiv F_{\mu\nu}F^{\mu\nu}$ comes from the Maxwell field,  $K\left(\phi\right)$ is the coupling function between scalar field and Maxwell field.  $V\left(\phi\right)$ is the scalar potential. When $K=1$ and $V=2\lambda$, it is the Einstein-Maxwell theory with cosmological constant $\lambda$. The theory gives the Reissner-Nordstrom-de Sitter solution. When $K=e^{2\phi}$ and $V=0$, it gives the dilaton black hole solution \cite{gib:1988,gar:1991}. When $K=e^{2\phi}$ and
\begin{equation}\label{eq:pot0}
V=\frac{\lambda}{3}\left(e^{2\phi}+4+e^{-2\phi}\right)\;,
\end{equation}
it gives the dilaton black hole in de Sitter universe \cite{gao:2004}. Except for these solutions, there are other important solutions with different $K$ and $V$ \cite{gao:2005a,Hajkhalili:2019,Hendi:2010,Hajkhalili:2018,Sheykhi2008,Yamazaki2001,anti}.  In principle, once the expressions of $K\left(\phi\right)$ and $V\left(\phi\right)$
are given, the gravity theory and the corresponding black hole solution are {determined}. This is the {conventional method}. But in this paper, we shall presume not $K\left(\phi\right)$ and $V\left(\phi\right)$, but the black hole solution in advance. Then $K\left(\phi\right)$ and $V\left(\phi\right)$ are subsequently {derived}.  {This is the solution-generating method \cite{cadoni:2011,Stephani:book}.} We require that the parameters such as the mass $M$ and electric charge $Q$ don't appear in the coupling function $K\left(\phi\right)$ and the scalar potential $V(\phi)$.

Now we derive the equations of motion. Varying the action with respect to the metric, Maxwell and dilaton fields, respectively, yields
\begin{eqnarray}
R_{\mu\nu}=2\nabla_{\mu}\phi\nabla_{\nu}\phi+\frac{1}{2}g_{\mu\nu}V+2K\left(F_{\mu\alpha}F^{\alpha}_{\nu}-\frac{1}{4}g_{\mu\nu}F^2\right)\;,
\end{eqnarray}
\begin{eqnarray}
\nabla_{\mu}\left(K F^{\mu\nu}\right)=0\;,
\end{eqnarray}
\begin{eqnarray}
\nabla_{\mu}\nabla^{\mu}\phi-\frac{1}{4}\left(V_{,\phi}+K_{,\phi}F^2\right)=0\;,
\end{eqnarray}
where ``$,\phi$'' denotes the derivative with respect to $\phi$.
The metric for static and spherically symmetric black hole solution can always be written as
\begin{equation}
ds^2=-U\left(r\right)dt^2+{U\left(r\right)}^{-1}dr^2+f\left(r\right)\left(d\theta^2+\sin^2\theta d\phi^2\right)\;.
\end{equation}
In this spacetime, the non-vanishing components of four-vector $A_{\mu}$ is uniquely $A_0(r)$.
So the equations of motion turn out to be
\begin{equation}\label{eom:1}
2ff^{''}+4f^2\phi^{'2}-f^{'2}=0\;,
\end{equation}
\begin{equation}\label{eom:2}
\left(fKA_0^{'}\right)^{'}=0\;,
\end{equation}
\begin{eqnarray}\label{eom:3}
&& 2f^2U^{''}+2fUf^{''}+2fU^{'}f^{'}-Uf^{'2}+4Uf^2\phi^{'2}\nonumber\\&&-4f^2KA_0^{'2}+2f^2V=0\;,
\end{eqnarray}
\begin{equation}\label{eom:4}
fU\phi^{''}+Uf^{'}\phi^{'}+fU^{'}\phi^{'}+\frac{1}{2}fK_{,\phi}A_0^{'2}-\frac{1}{4}fV_{,\phi}=0\;.
\end{equation}
Here prime denotes the derivative with respect to $r$. Observing these equations, we find six unknown functions. They are $U,\ f,\ \phi,\ A_0,\ K,\ V$. But we have only four equations of motion. Therefore, the equations of motion are not closed. In principle, one can presume any two of them in advance. Then the equations of motion are closed. In the next section, we {shall employ the solution-generating techniques \cite{cadoni:2011,Stephani:book} to construct new black holes }.

%shall bring forward $U$ and $f$ beforehand. Afterwards, the other four functions are worked out. By this way, we construct two black hole solutions.%

\section{The first solution}
We recall that when $K=e^{2\phi}$ and $V=0$, the theory gives the dilaton black hole solution
with \cite{gib:1988,gar:1991}
\begin{eqnarray}
U&=&1-\frac{2M}{r}\;,\ \ \ \ f=r\left(r-\frac{Q^2}{M}\right)\;,\nonumber\\
\phi &=&-\frac{1}{2}\ln\left(1-\frac{Q^2}{M{r}}\right)\;.
\end{eqnarray}
Here $M$ and $Q$ have the physical meaning of ADM mass and electric charge of the black hole, respectively. This spacetime is asymptotically flat. On the other hand,
when $K=e^{2\phi}$ and $V$ of Eq.~(\ref{eq:pot0}), the theory gives the dilaton black hole in de Sitter universe with \cite{gao:2004}
\begin{eqnarray}\label{eq:gao}
U&=&1-\frac{2M}{r}-\frac{1}{3}\lambda f\;,\ \ \ \ f=r\left(r-\frac{Q^2}{M}\right)\;,\nonumber\\
\phi &=&-\frac{1}{2}\ln\left(1-\frac{Q^2}{M{r}}\right)\;.
\end{eqnarray}
Here $\lambda$ is a constant. When $Q=0$, it goes back  to the Schwarzschild-de Sitter solution. In order to construct a new solution, we observe the expression of $U$ and find that the $\lambda$ term is {not} proportional to $r^2$, but $f$. Inspired by this fact, we presume the new solution
\begin{equation}\label{eq:uf1}
U=1-\frac{2M}{r}+\frac{\beta Q^2}{f}-\frac{1}{3}\lambda f\;,
\end{equation}
\begin{equation}\label{eq:uf2}
f=r\left(r+\gamma\frac{Q^2}{M}\right)\;.
\end{equation}
Here $\beta$ and $\gamma$ are dimensionless constants. The $\beta$
term is a new term and it is similar to the $Q^2/r^2$ term in Reissner-Nordstrom-de Sitter solution.
Then substituting equations (\ref{eq:uf1}) and (\ref{eq:uf2}) into the equations of motion (\ref{eom:1},\ref{eom:2},\ref{eom:3},\ref{eom:4}), we obtain
\begin{equation}
\phi=-\frac{1}{2}\ln\left(1+\frac{\gamma Q^2}{Mr}\right)\;,
\end{equation}
\begin{equation}
A_0=\frac{\gamma Q}{r}-\frac{\beta Q}{2r}-\frac{\beta M Q}{2\left(Mr+\gamma Q^2\right)}\;,
\end{equation}
\begin{equation}
K=\frac{2e^{2\phi}}{-2\gamma+\beta+\beta e^{4\phi}}\;,
\end{equation}
\begin{equation}\label{eq:pot1}
V=\frac{\lambda}{3}\left(e^{2\phi}+4+e^{-2\phi}\right)\;.
\end{equation}
We notice that the scalar potential, Eq.~(\ref{eq:pot1}) is exactly that in Eq.~(\ref{eq:pot0}). Two dimensionless constants $\gamma$ and $\beta$ appear in the coupling function $K$. Up to this point, equations from Eq.~(\ref{eq:uf1}) to Eq.~(\ref{eq:pot1}) constitute the first black hole solution in this paper.
When $\gamma=-1,\ \beta=0$, the solution returns to Eq.~(\ref{eq:gao}).

It seems there are two dimensionless constants $\gamma$ and $\beta$ in the coupling function $K$. We find this not the case. Actually, if we rescale $A_{\mu}\rightarrow A_{\mu}\sqrt{-\gamma}$ and $\beta\rightarrow-\beta\gamma$, the coupling constant $\gamma$ disappears (or is gauged to minus one). Then we are left with unique coupling constant $\beta$. Therefore, in the next, we adopt $\gamma=-1$. In TABLE.~I, we list the three solutions for {comparison}.
When $\beta=0$, the solution returns to Eq.~(\ref{eq:gao}).

To understand the role of coupling constant $\beta$, we plot the evolution of $K$ with respect to $\phi$ for different $\beta$ in Fig.~{\ref{FigK}}. Fig.~{\ref{FigK}} shows that with the increasing of $\beta$, the effect of Maxwell invariant becomes smaller and smaller. When $\beta\rightarrow +\infty$, gravity takes over electromagnetic interaction and the electromagnetic field can be safely neglected. This means the value of $\beta$ determines the strength of interaction between Maxwell field and the scalar field. The solution, as given by Eq.~(15)-(18), is purely electric. We remember that in standard dilaton theory there is an electric-magnetic duality transformation
\begin{equation}
\{P\rightarrow Q,\ \ Q\rightarrow P\} \ \ \textrm{and}\ \ \ \ {K\left(\phi\right)\rightarrow\frac{1}{K\left(\phi\right)}}\;,\nonumber\\
\end{equation}
which turns the electric field into a magnetic field and changes the sign of $\phi$ without modifying the spacetime metric. We find this does not apply to the solution except for $\beta=0$ and $\beta=2\gamma$.

Now let's consider {the} situation of $\lambda=0$ for the moment. We can make a translation transformation $\phi\rightarrow \phi+\phi_0$ with $\phi_0$ a constant on the coupling function $K(\phi)$ such that $K(\phi)\propto 1/\cosh{2\phi}$. Then there is a maximum for $K(\phi)$ at $\phi=0$. At the maximum,  we have $dK/d\phi=0$. In this case, $\phi=0$ solves the field equations and the Reissner-Nordstrom (RN) spacetime is a solution. However, the RN solution is not unique to the theory. As we show, Eq.~(13)-(17) constitute the second set of black hole solutions, with the nontrivial scalar field profile given by Eq.~(15). It is the corresponding scalarised counterparts of RN black holes.

The effective mass squared of the scalar field ($\lambda=0$) is
\begin{equation}
\mu^2_{\textrm{eff}}=\frac{F^2}{4}\frac{d^2 K}{d\phi^2}|_{\phi=0}>0\;,
\end{equation}
 because we have $F^2<0$ for the pure electric field. Therefore, there is no the tachyonic instability. In other words, the RN solution is stable to the scalar perturbations. According to the classification of Ref.~\cite{asto:2019}, this scalarised solution belongs to the Subclass IIB or scalarised-disconnected-type. In this case, the scalarised black holes do not bifurcate from RN black holes, and do not continuously reduce to the latter when $\phi=0$.

\begin{table}[]
\begin{center}
\begin{tabular}{|c|c|c|c|c|c|c|}
\hline \hline
 &$U(r)$& $f(r)$&$K(r)$&$V(r)$&$\phi(r)$& $A(r)$\\\hline
Dilaton &$1-\frac{2M}{r}$&$r^{2}(1-\frac{Q^{2}}{Mr})$&$e^{2\phi}$&$0$&$-\frac{1}{2}\text{ln}(1-\frac{Q^{2}}{Mr})$& $-\frac{Q}{r}$\\\hline
Dilaton&&&&&&\\
 de-Sitter&$1-\frac{2M}{r}-\frac{1}{3}\lambda f$&$r^{2}(1-\frac{Q^{2}}{Mr})$&$e^{2\phi}$&$\frac{\lambda}{3}\left(e^{2\phi}+4+e^{-2\phi}\right)$&$-\frac{1}{2}\text{ln}(1-\frac{Q^{2}}{Mr})$& $-\frac{Q}{r}$\\\hline
First&&&&&&\\
solution&$1-\frac{2M}{r}+\frac{\beta Q^{2}}{f}-\frac{1}{3}\lambda f$&$r^{2}(1-\frac{Q^{2}}{Mr})$&$\frac{2e^{2\phi}}{\beta e^{4\phi}+\beta+2}$&$\frac{\lambda}{3}\left(e^{2\phi}+4+e^{-2\phi}\right)$&$-\frac{1}{2}\text{ln}(1-\frac{Q^{2}}{Mr})$& $-\frac{Q}{r}-\frac{\beta Q}{2}\left(\frac{1}{r}+\frac{r}{f}\right)$\\
\hline \hline
\end{tabular}
\label{table111}
\end{center}
\end{table}

The condition of $dK/d\phi|_{\phi=0}=0$ guarantees the existence of RN solution, but not the general scalarised black hole solutions. In order to guarantee there are scalarised black hole solutions, two Bekenstein-type identities should be satisfied \cite{asto:2019}.

The first identity is given by
\begin{equation}
\int\sqrt{-g}d^4x\left(K_{\phi\phi}\nabla_{\mu}\phi\nabla^{\mu}\phi+\frac{K_{\phi}^2}{4}F^2\right)=0\;.
\end{equation}
For a purely electric field, one has $F^2<0$. This implies
\begin{equation}
K_{,\phi\phi}>0\;,
\end{equation}
should be satisfied in some region of $r$. Otherwise, the two terms of the integrand
 will have always the same sign which makes the identity holds if and only if $\phi=0$.

The second identity is given by
\begin{equation}
\int\sqrt{-g}d^4x\left(\nabla_{\mu}\phi\nabla^{\mu}\phi+\frac{{\phi K_{,\phi}}}{4}F^2\right)=0\;.
\end{equation}
This reveals that for a pure electric field, the potential should satisfy the condition
\begin{equation}
\phi K_{,\phi}>0\;,
\end{equation}
in some region of $r$. In Fig.~\ref{fig:bek}, we present the plots of $K_{,\phi\phi}$ and $\phi K_{,\phi}$ with $K=1/\cosh{2\phi}$ and $Q^2/M=0.5$. It is apparent we always have $K_{,\phi\phi}>0$ and $\phi K_{,\phi}>0$ in some range of $r$. Then we conclude that the theory has the scalarised black hole solution with pure electric field. The conclusion is in agreement with the solution given by Eq.~(15-17) for $\lambda=0$.

\begin{figure}[h]\label{FigK}
\begin{center}
\includegraphics[width=9cm]{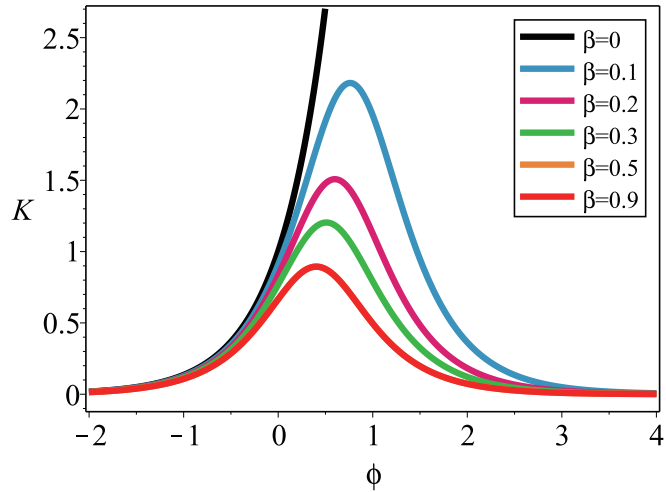}
\caption{The evolution of coupling function $K(\phi)$. $K(\phi)$ is regular in the entire field space. The plots correspond to $\beta=0,\ \ 0.1,\ \ 0.2,\ \ 0.3,\ \ 0.5,\ \ 0.9$, respectively, from top down.}\label{FigK}
\end{center}
\end{figure}

\begin{figure}[h]
\begin{center}
\includegraphics[width=9cm]{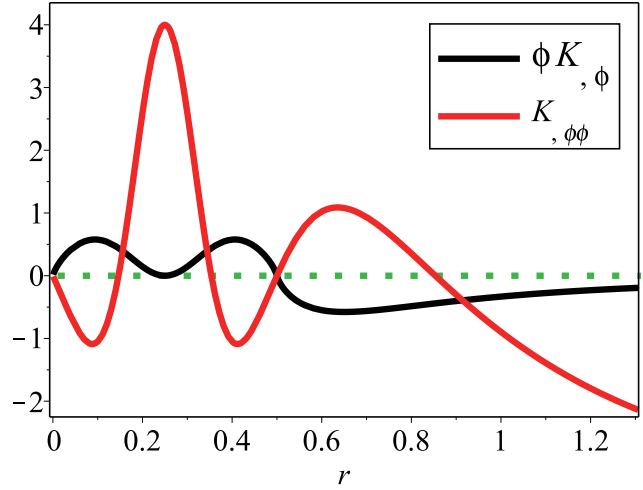}
\caption{The plots of $K_{,\phi\phi}$ and $\phi K_{,\phi}$ ($K=1/cosh{2\phi}$) with respect to $r$, respectively. In some region of $r$, one always have $K_{,\phi\phi}>0$ and $\phi K_{,\phi}>0$. Therefore, the theory has the scalarised black hole solution with pure electric field.}\label{fig:bek}
\end{center}
\end{figure}

In the following, we shall compute the scalar charge $D$, the electric charge $Q$ and the ADM mass $M$ of the black hole.
The scalar charge $D$ can be computed by
\begin{equation}
D=\frac{1}{4\pi}\oint \nabla_{\mu}\phi \ dS^{\mu}=-\frac{Q^2}{2M}\;.
\end{equation}
Here the integration is taken over a spacelike surface enclosing the origin. The scalar charge is a conserved value and thus it does not depend on the choice of the surface. It is clear that the scalar charge is determined by the mass $M$ {and} electric charge $Q$.

The electric charge of the black hole is shown to be
\begin{equation}\label{charge}
Q=\frac{1}{4\pi}\oint K \left({\phi}\right)\nabla_{\mu}A_{0}\ dS^{\mu}\;.
\end{equation}
The electric charge is also a conserved value and it does not depend on the choice of the surface.

\begin{figure}[h]
\centering
\subfigure[]{\label{fig:ur}\includegraphics[width=0.47\textwidth]{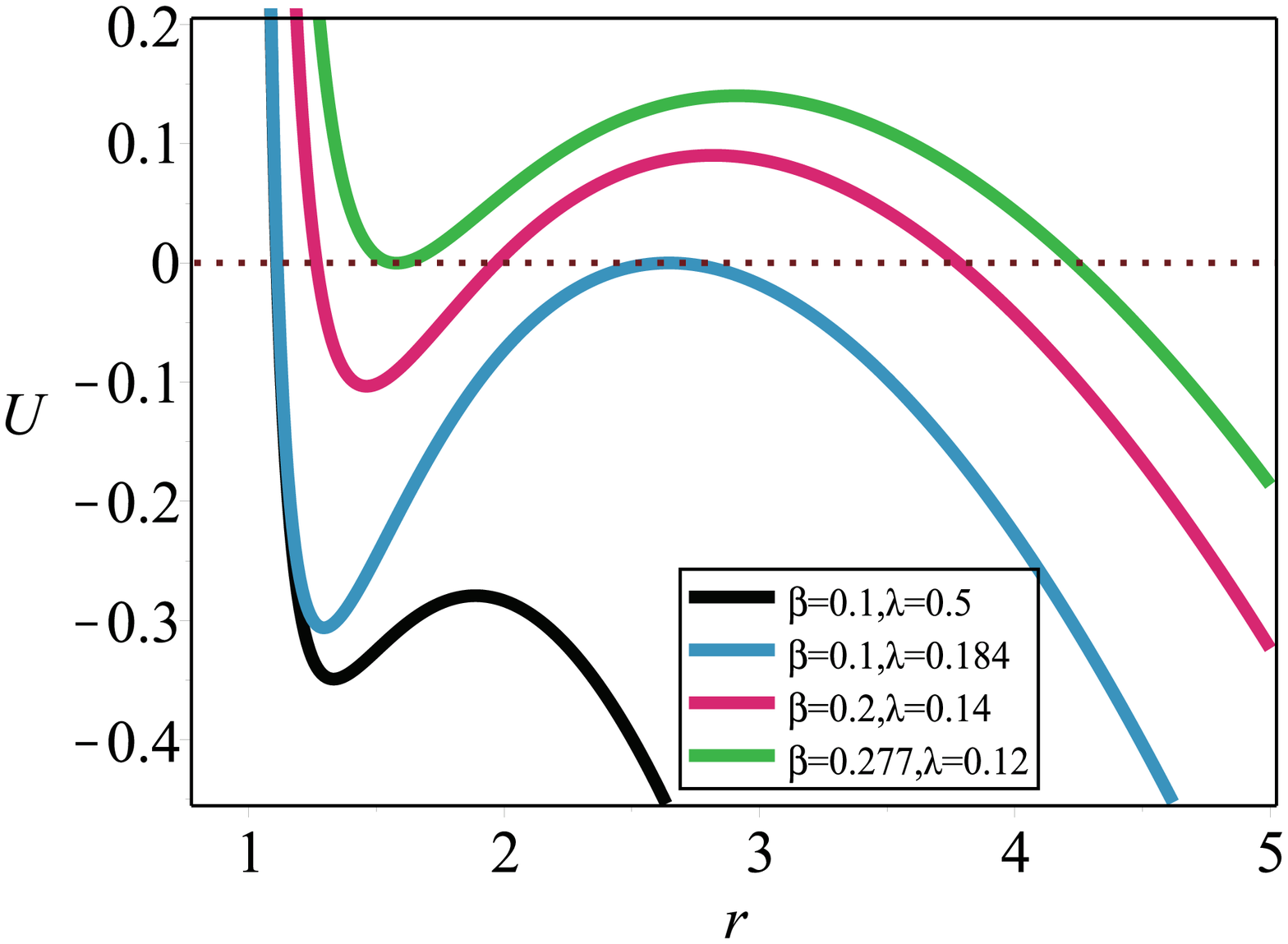}}
\subfigure[]{\label{fig:urp0}\includegraphics[width=0.47\textwidth]{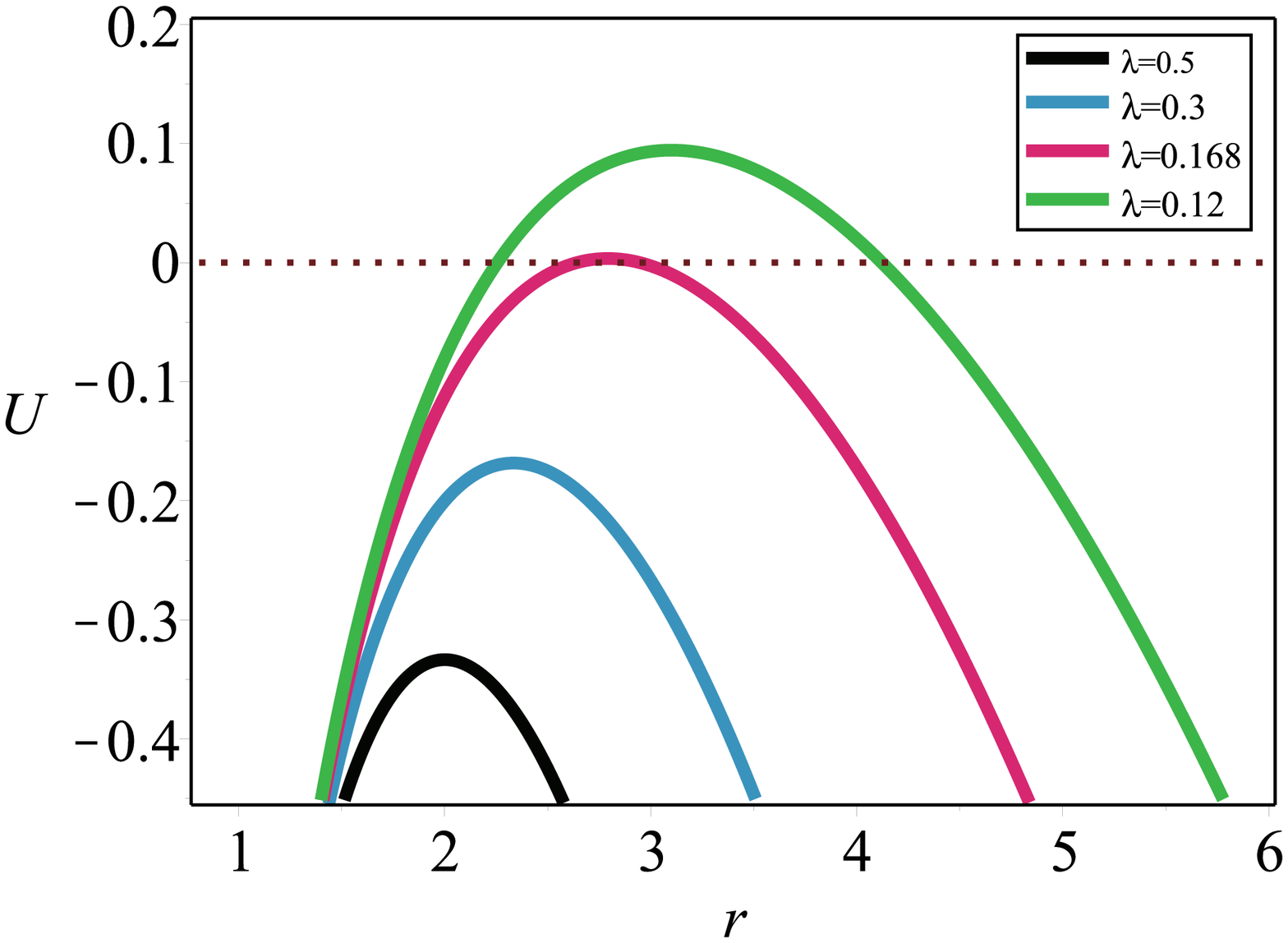}}
\subfigure[]{\label{fig:urn}\includegraphics[width=0.47\textwidth]{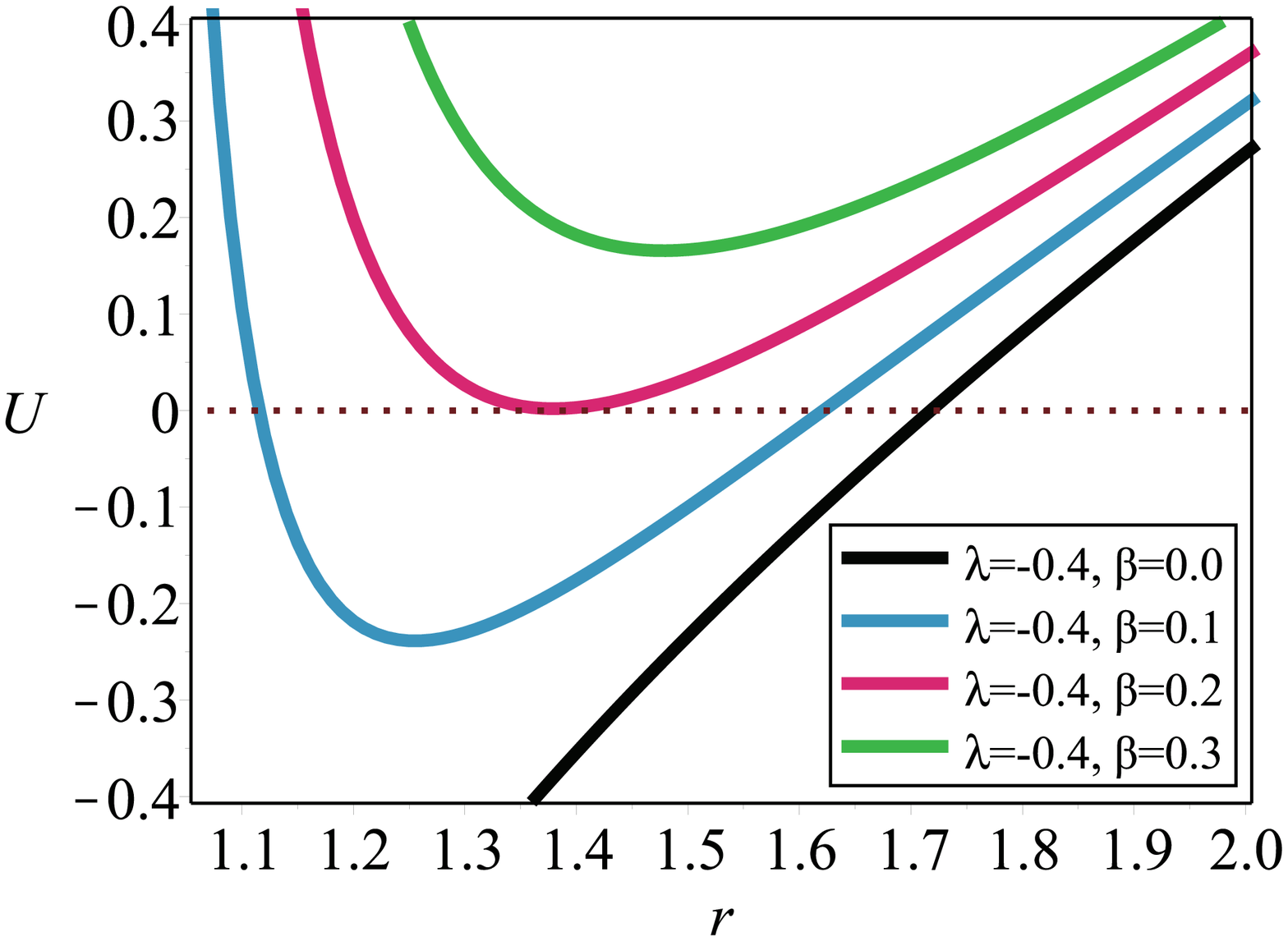}}
\caption{{Variation of $U$ with respect to} $r$. (a) There are four cases for the scenario of horizons. The curves correspond to 1-horizon, 2-horizon, 3-horizon and 2-horizon from down to up, respectively. We fix $M=1,\ Q=1$ and choose $\beta=0.1,\ \lambda=0.5$; $\beta=0.1,\ \lambda=0.184$; $\beta=0.2,\ \lambda=0.14$ and $\beta=0.277,\ \lambda=0.12$, respectively. (b) There are three cases for the scenario of horizons {with vanishing} $\beta$ term. The curves correspond to 0-horizon, 0-horizon, 1-horizon and 2-horizon from down to up, respectively. We fix $M=1,\ Q=1$ and choose $\lambda=0.5$, $ \lambda=0.3$, $\lambda=0.168$ and $\lambda=0.12$, respectively. (c) There are three cases for the scenario of horizons. The curves correspond to 0-horizon, 1-horizon, 2-horizon and 1-horizon from up to down, respectively. We fix $M=1,\ Q=1,\ \lambda=-0.4$ and choose $\beta=0.0\ $, $\beta=0.1\ $, $\beta=0.2\ $ and $\beta=0.3$, respectively.  }
\end{figure}

The ADM mass \cite{ADM:1,ADM:2,ADM:3} satisfies
\begin{equation}
M=\frac{1}{16\pi}\lim_{S^{\mu}\rightarrow i^{0}}\oint g^{\alpha\beta}\left(g_{\alpha\mu,\beta}-g_{\alpha\beta,\mu}\right)dS^{\mu}\;,
\end{equation}
where $i^{0}$ is the spacetlike infinity. The Ricci scalar $R$ and the Maxwell invariant $F_{\mu\nu}F^{\mu\nu}$ of the spacetime are give by
\begin{eqnarray}
R&=&-\frac{4 f^{''} U f+2U^{''} f^{2}-f^{'2} U+4U^{'}f^{'} f-4 f}{2f^{2}}\\\;
&=&\frac{2}{r^{3}\left(r+2D\right)^{3}}\left(2 \lambda r^{6}+12 D \lambda r^{5}+25 D^{2} \lambda r^{4}+20 D^{3} \lambda r^{3}\right. \;, \nonumber\\&&
+\left.\left(4 D^{4} \lambda+D^{2}\right) r^{2}+\left(2 D^{3}-2 D^{2} M\right) r+D^{2} Q^{2} \beta-4 D^{3} M\right)\;,
\end{eqnarray}
%\begin{eqnarray}
%R&=&-\frac{1}{2M^2r^4\left(r+2D\right)}\left[8\lambda M^3 r^6+\lambda\gamma^3Q^6r^3-3\beta M\gamma^2Q^6-2M^2r\gamma^2 Q^4+\gamma^2Q^4r^2M\right.\nonumber\\&&\left.+9r^4 M\lambda\gamma^2Q^4+16r^5M^2\lambda\gamma Q^2-4rM^2\beta Q^4\gamma\right]\;,
%\end{eqnarray}

\begin{eqnarray}\label{negative}
F_{\mu\nu}F^{\mu\nu}&=&-2 A0^{'2}=-2\left[\frac{ Q}{r^2}+\frac{\beta Q}{2 r^2}+\frac{\beta Q}{2\left(r+2D\right)^2}\right]^2\;.
\end{eqnarray}
%\begin{eqnarray}
%F_{\mu\nu}F^{\mu\nu}&=&-2\left[-\frac{\gamma Q}{r^2}+\frac{\beta Q}{2 r^2}+\frac{\beta Q }{2\left(r+2D\right)^2}\right]^2\;.
%\end{eqnarray}
From the definition of $f$ in Eq.~({\ref{eq:uf2}}) we know $r$ should satisfy
\begin{eqnarray}
r\geq-2D>0\;.
\end{eqnarray}
We conclude {that} both the Ricci scalar and the Maxwell invariant are divergent at $r=-2D$. Therefore, the position of $r=-2D$ is the physical singularity of this spacetime. As for the black hole horizons, we find that, under the condition {of} $\beta>0$, there are at most three horizons for positive $\lambda$. They are cosmic horizon, black hole event horizon and black hole inner horizon, respectively, as shown in Fig.~\ref{fig:ur}. For negative $\lambda$, there are at most two horizons. They are black hole event horizon and black hole inner horizon. As a result, in the presence of cosmological constant $\lambda$, the causal structure (Penrose diagram) of the black hole spacetime is exactly the same as the Reissner-Nordstrom-de Sitter (or anti-de Sitter) spacetime.

%%\blue{Two further parameters $\beta$ and $\gamma$ can affect black hole horizon.(Fig.~\ref{fig:ur}) In fact, $\beta$ term can affect the exist of inner horizon(Fig.~\ref{fig:urp0}). If $\beta$ term is missing, we can know that the inner horizon vanished as well. If we take the series expansion of the horizon equation, we can see the $\beta$ parameter can describe coupling strength.}

We find that it is rather involved to study the variation of horizons in the presence of cosmological constant. So here we shall focus on the case of $\lambda=0$. In order to analyse the {scenario} of horizons, we make coordinate transformation as follows
\begin{eqnarray}
r=\frac{Q^2+\sqrt{Q^4+4M^2 x^2}}{2M}\;.
\end{eqnarray}
Then the metric is presented in the Schwarzschild coordinate system
\begin{eqnarray}
ds^2-Udt^2+\frac{4M^2x^2}{U\left(Q^4+4M^2x^2\right)}dx^2+x^2d \Omega^2\;.
\end{eqnarray}
with
\begin{eqnarray}
U=1-\frac{4M^2}{Q^2+\sqrt{Q^4+4M^2x^2}}+\frac{\beta Q^2}{x^2}\;.
\end{eqnarray}
Now the physical singularity is at $x=0$ and the positions of horizons are obtained by solving $U=0$,
\begin{eqnarray}
x_{+}=\sqrt{-Q^2-\beta Q^2+2M^2+\sqrt{Q^4+4M^4-4Q^2M^2-4\beta Q^2M^2}}\;,
\end{eqnarray}
\begin{eqnarray}
x_{-}=\sqrt{-Q^2-\beta Q^2+2M^2-\sqrt{Q^4+4M^4-4Q^2M^2-4\beta Q^2M^2}}\;.
\end{eqnarray}
$x_+$ and $x_-$ represent the event horizon and the inner Cauchy horizon, respectively.
Defining the ratio $\epsilon$ of charge to mass {as} $\epsilon=Q/M$, we have

\begin{eqnarray}
x_{+}=M\sqrt{2-\epsilon^2-\beta\epsilon^2+\sqrt{\left(2-\epsilon^2\right)^2-4\epsilon^2\beta}}\;,
\end{eqnarray}
\begin{eqnarray}
x_{-}=M\sqrt{2-\epsilon^2-\beta\epsilon^2-\sqrt{\left(2-\epsilon^2\right)^2-4\epsilon^2\beta}}\;.
\end{eqnarray}
Without the loss of generality, we assume $Q>0$. Then we have the following conclusions.

(1) When
\begin{eqnarray}
\epsilon\geq \sqrt{2}\;,\ \ \ \beta>0\;,
\end{eqnarray}
there are no horizons and the central singularity is naked. This reveals the maximum charge of the black hole should be smaller than $\sqrt{2}M$ in order that the singularity is dressed with horizon.

(2) When
\begin{eqnarray}
\epsilon=0\;,
\end{eqnarray}
it reduces to the {Schwarzschild} solution.

(3) When \begin{eqnarray}
0<\epsilon<\sqrt{2}\;,\ \  \textrm{and} \  \  \  \beta>\frac{\left(2-\epsilon^2\right)^2}{4\epsilon^2}\;,
\end{eqnarray}
there are no horizons and the central singularity is naked. This reveals that the singularity can be naked
if the coupling constant $\beta$ is sufficiently large no matter how small the charge is.

(4) When \begin{eqnarray}
0<\epsilon<\sqrt{2}\;,\ \  \textrm{and} \  \  \  \beta<\frac{\left(2-\epsilon^2\right)^2}{4\epsilon^2}\;,
\end{eqnarray}
there are two horizons.

(5) When \begin{eqnarray}
0<\epsilon<\sqrt{2}\;,\ \  \textrm{and} \  \  \  \beta=\frac{\left(2-\epsilon^2\right)^2}{4\epsilon^2}\;,
\end{eqnarray}
the two horizons coincide and the so-called extreme black hole is achieved. This tells us the black hole can be extreme
if the coupling constant $\beta$ {meets above condition} no matter how small the charge is.

In the next section, we shall study the thermodynamics of this spacetime.

\section{black hole thermodynamics}

In order to explore the black holes thermodynamics, we start from the calculation of its temperature. The black hole temperature is well defined in quite general setting due to its strict geometrical basis \cite{Gib:1977}. One can employ the so-called surface gravity which is defined as follows:
\begin{equation}
\kappa^2=-\frac{1}{2}\nabla_{\mu}\chi_{\nu}\nabla^{\mu}\chi^{\nu}\;,
\end{equation}
where $\chi_{\mu}$ is a Killing vector field and it is null on the horizon. We can chose $\chi^{\nu}=\partial/\partial t$ because spacetime is the static. As a result, the black hole temperature takes the form
\begin{eqnarray}\label{temperature}
T&=&\frac{{\kappa}}{2\pi}=\frac{1}{4\pi}U^{'}\left(r_{+}\right)\;.
\end{eqnarray}

\begin{figure}[htbp]
\centering
\subfigure[]{  \label{Fig2f}
\includegraphics[width=8cm,height=6cm]{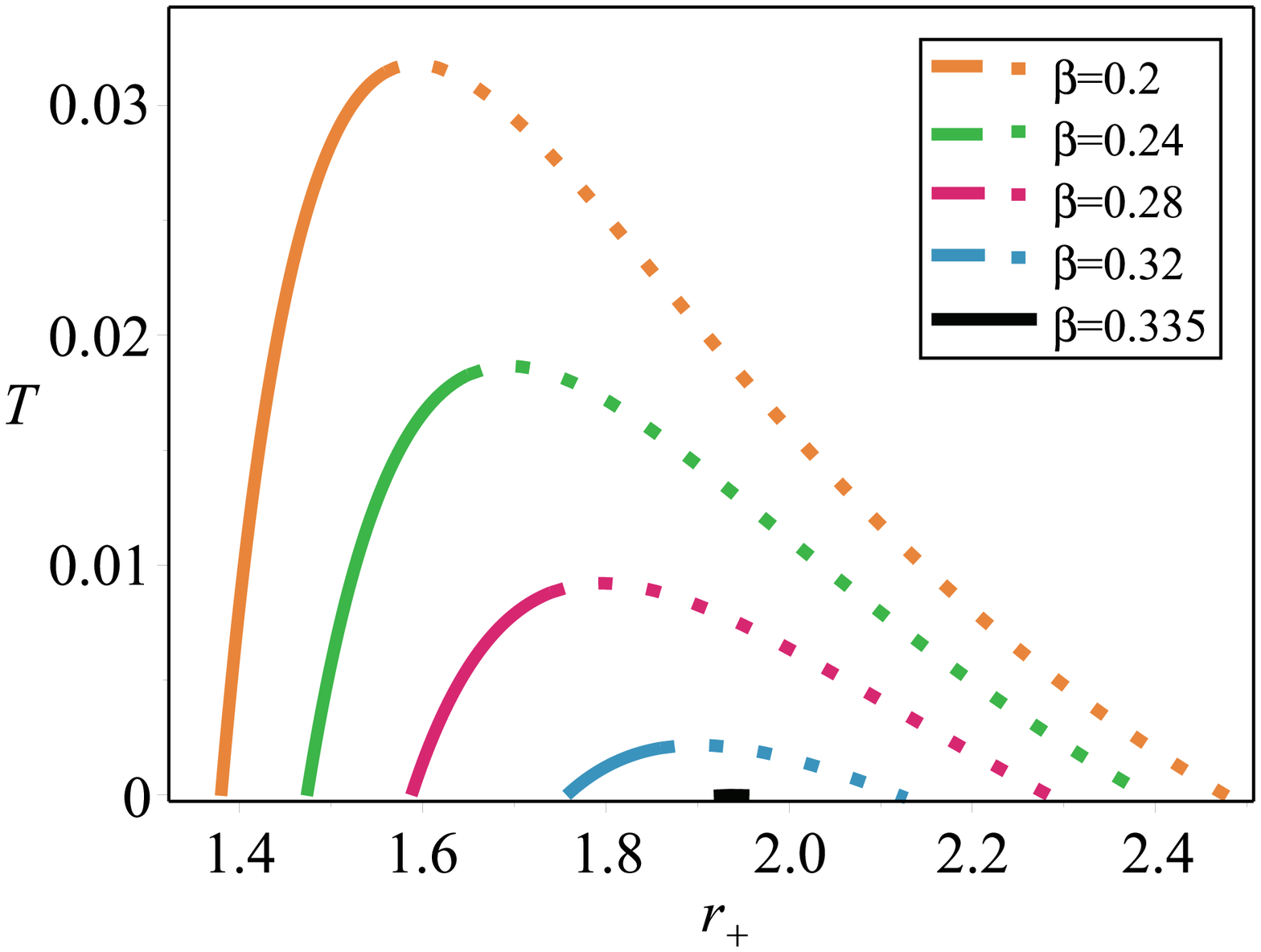}}\subfigure[]{\label{Fig4f}
\includegraphics[width=8cm,height=6cm]{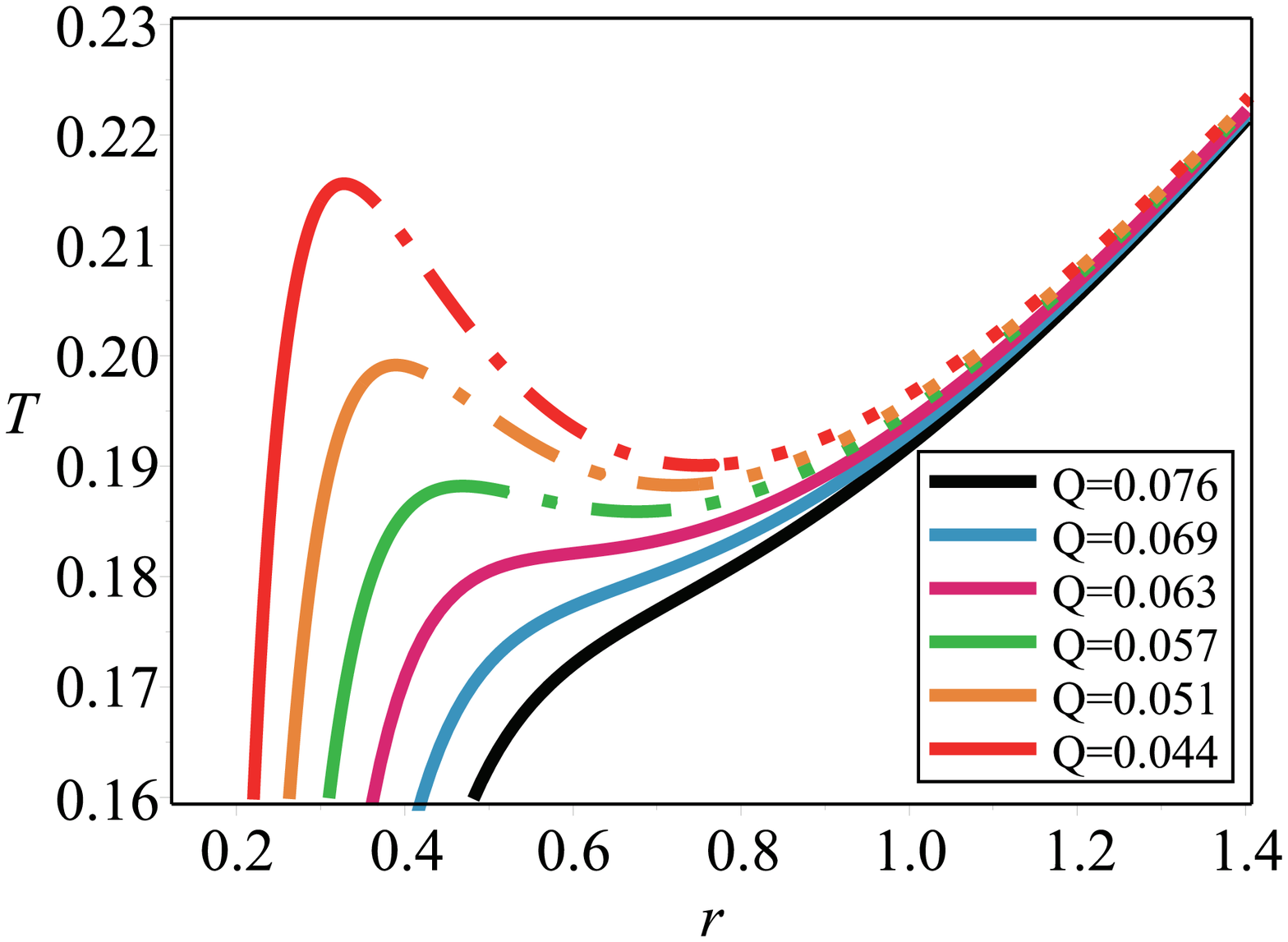}}\\
\caption{Black hole temperature $T$ as the function of black hole event horizon $r_+$. (a) The curves correspond to $\beta=0.2,\ 0.24,\ 0.28,\ 0.32,\ 0.335$, from up to down, respectively. We put  $M=1,\ Q=1$.  There are two phases of black holes with the same temperature, the so-called small (solid lines) and large black holes (dotted lines), respectively. When $\beta=0.335$, there is uniquely one phase and its temperature is zero. (b) The curves correspond to $Q=0.076,\ 0.069,\ 0.063,\ 0.057,\ 0.051,\ 0.044$, from down to up, respectively. We put  $\lambda=-1.5,\ \beta=0.14$.  With the decreasing of $Q$, the black holes make phases transition from 1-phase to 3-phase which consists of small (solid line), middle (dash-dot line) and large (dotted line) black holes.}
\end{figure}

Substituting the first black hole solution metric Eq.(\ref{eq:uf1}), we get
\begin{eqnarray}\label{eq:t1}
T&=&\frac{M}{2\pi r_+^2}-\frac{\beta Q^2}{4\pi r_+^2\left(r_++2D\right)}-\frac{\beta Q^2}{4\pi r_+\left(r_++2D\right)^2}\nonumber\\&&-\frac{\lambda r_{+}}{6\pi}+\frac{\lambda Q^2}{12\pi M}\;,
\end{eqnarray}
where $r_+$ represents the black hole event horizon which is determined by $U\left(r_+\right)=0$. In Fig.~\ref{Fig2f} and Fig.~\ref{Fig4f} we plot the black hole temperature $T$ as the function of black hole event horizon $r_+$ with running $\beta$ and $Q$, respectively. In Fig.~{4a},  we put  $M=1,\ Q=1$ and $\beta=0.2,\ 0.24,\ 0.28,\ 0.32,\ 0.335$. Given $M,\ Q,\ \beta$, then $\lambda$ can be expressed as the function of $\lambda(M,Q,\beta,r_{+})$ by using the event horizon equation. Eventually, the temperature $T$ becomes the parametric equations of $r_{+}$. Fig.~\ref{Fig2f} shows there are two phases of black holes with the same temperature, the so-called small and large black holes, respectively. When $\beta=0.335$, there is uniquely one phase and its temperature is zero. With the increasing of $\beta$, the black holes make phases transition from 2-phase to 1-phase. In Fig. 4(b), we put  $\lambda=-1.5,\ \beta=0.14$ and  $Q=0.069,\ 0.063,\ 0.057,\ 0.051,\ 0.044$. Given $\lambda,\ Q,\ \beta$, then $M$ can be expressed as the function of $M(\lambda,Q,\beta,r_{+})$ by using the event horizon equation. Eventually,  temperature $T$ becomes the parametric equations of $r_{+}$. Fig.~\ref{Fig4f} shows that there are three phases of black holes with the same temperature. They are large, middle and small black holes, respectively. With the decreasing of $Q$, the black holes make phases transition from 1-phase to 3-phase.

The entropy of black holes generally satisfies the area law which states that the entropy is a quarter of the area of black hole event horizon \cite{beck:1973}.
The law applies to almost all kinds of black holes including Einstein-Maxwell-scalar black holes \cite{haw:1999}. Therefore we have the entropy of the black hole
 \begin{equation}
S=\frac{A}{4}=\pi r_{+}\left(r_{+}+2D\right)\;.
\end{equation}

According to definition of \cite{cve:1999}, the electric potential measured by the distant observer is

\begin{equation}\label{eq:p1}
\Phi=\int_{r_{+}}^{+\infty}A_0^{'}dr=-A_0\left(r_{+}\right)\;.
\end{equation}

The study on the thermodynamic phase structure of AdS black holes \cite{kas:2009} tells us the cosmological constant acts as a thermodynamic pressure

\begin{equation}\label{eq:press}
P=-\frac{\lambda}{8\pi}\;.
\end{equation}

The variation of thermal pressure requires the presence of a conjugate thermal volume in the first law of thermodynamics, which can lead to a variety
of novel thermodynamic behaviour, for example, triple points \cite{alt:2014}, reentrant phase transitions
\cite{alt:2013}, the emergence of polymer-like  phase structure \cite{fra:2014}, the superfluid-like phase structure \cite{hen:2017} and
the Van der Waals transition \cite{kub:2012}.

We find the conjugate thermodynamic volume is

\begin{equation}\label{eq:v1}
\mathfrak{V}\equiv\left(\frac{\partial M}{\partial P}\right)_{S,Q}=\frac{4\pi}{3}r_{+}\left(r_{+}+2D\right)\left(r_{+}+D\right)\;.
\end{equation}
Then the Smarr formula
\begin{equation}\label{Smarr}
M=2TS-2\mathfrak{V}P+Q\Phi\;,
\end{equation}
is satisfied.

In the following, we shall make an examination whether the thermal quantities satisfy the requirement of the first law of thermodynamics. From the equation of horizon, we obtain
\begin{equation}\label{eq:lamb}
\lambda=\frac{3}{f_+}\left(1-\frac{2M}{r_+}+\frac{\beta Q^2}{f_+}\right)\;,
\end{equation}
where $f_+=f\left(r=r_+\right)$. So in view of Eq.~(\ref{eq:press}), the pressure $P$ can be written as

 \begin{equation}\label{eq:prs}
P=-\frac{3}{8\pi f_+}\left(1-\frac{2M}{r_+}+\frac{\beta Q^2}{f_+}\right)\;.
\end{equation}
We treat the pressure $P$, the entropy $S$ as the function of $r_+,\ M,\ Q$. Then we have
\begin{equation}
dP=P_{,M}dM+P_{,Q}dQ+P_{,r_{+}}dr_+\;,
\end{equation}

\begin{equation}
dS=S_{,M}dM+S_{,Q}dQ+S_{,r_{+}}dr_+\;.
\end{equation}
By using the equations Eq.~(\ref{eq:t1}), Eq.~(\ref{eq:p1}) and Eq.~(\ref{eq:v1}), we find the first law of thermodynamics
\begin{equation}\label{first law}
dM=TdS+\Phi dQ+\mathfrak{V}dP\;,
\end{equation}
is indeed satisfied.

It is pointed \cite{kas:2009,gun:2012} that once the thermodynamic pressure $P$ and thermal volume $\mathfrak{V}$ are introduced, the ADM mass $M$ should be understood as enthalpy. Then the more convenient function to analyze thermodynamic behaviour of a system, especially in the case that there is some critical behaviour, is the Gibbs free energy which is defined in the following way
\begin{equation}
G=M-TS\;.
\end{equation}
The Gibbs free energy $G$ is understood to depend on $M,\ Q,\ \beta,\ r_+$.

\begin{figure}[htbp]
\centering
\subfigure[]{  \label{Fig3f}
\includegraphics[width=8cm,height=6cm]{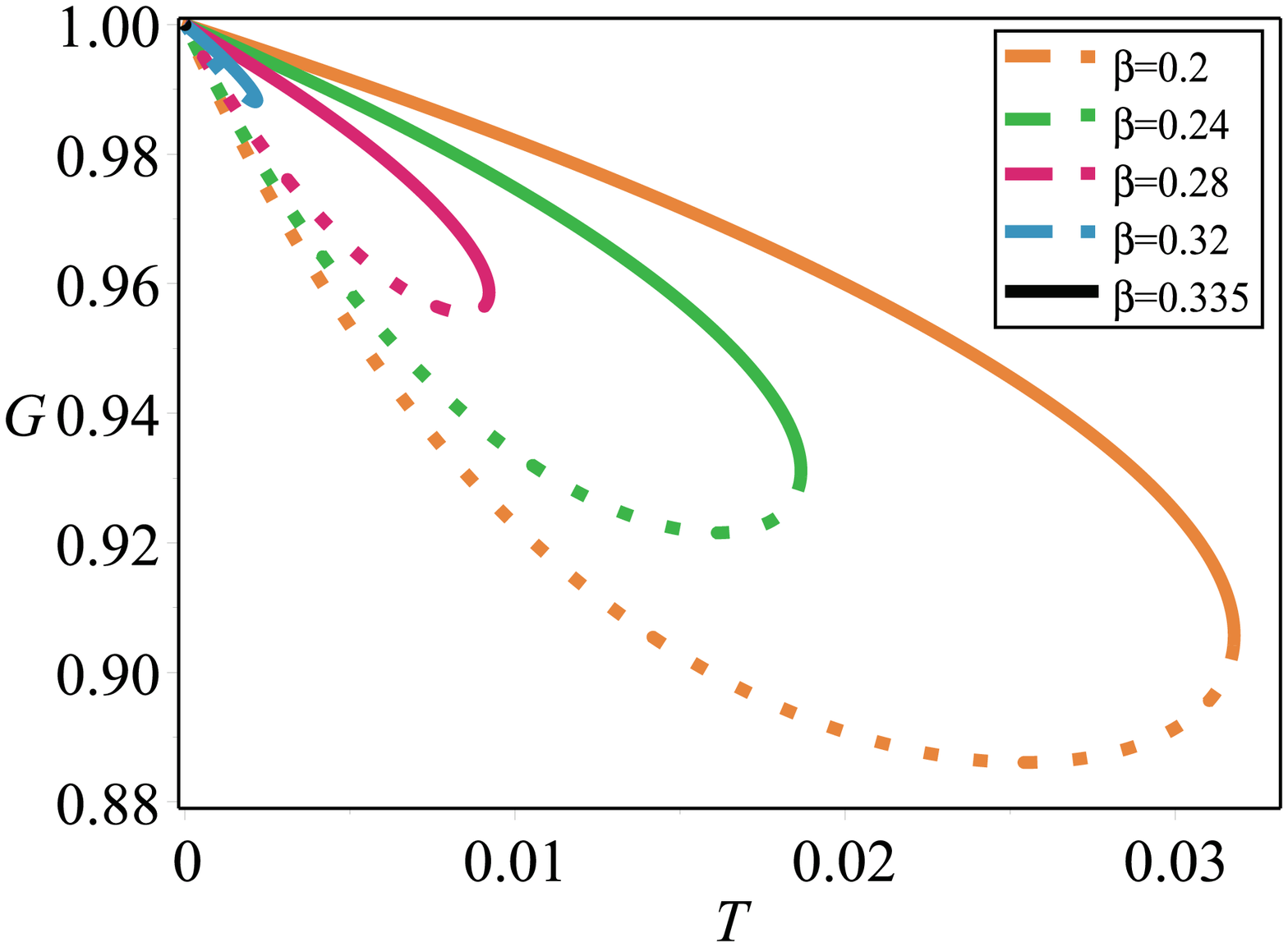}}\subfigure[]{\label{Fig5f}
\includegraphics[width=8cm,height=6cm]{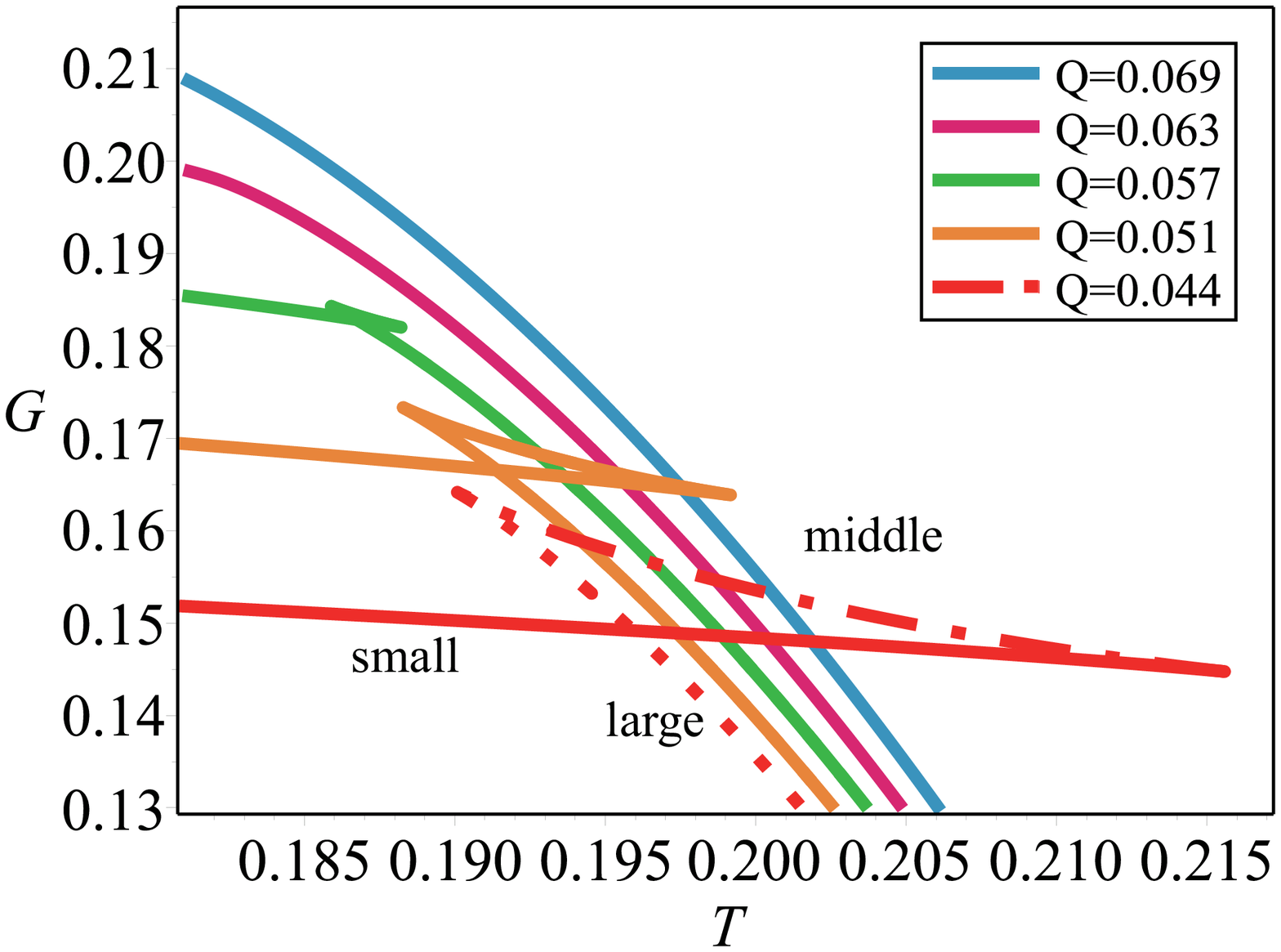}}\\
\caption{Black hole Gibbs free energy $G$ as the function of temperature $T$. (a) The curves correspond to $\beta=0.2,\ 0.24,\ 0.28,\ 0.32,\ 0.335$, from outer to inner, respectively. We put  $M=1,\ Q=1$.  There are two phases of black holes (dotted lines for small black hole and solid lines for large black holes) with the same temperature. When $\beta=0.335$, the Gibbs free energy is $G=M=1$. (b) The curves correspond to  $Q=0.069,\ 0.063,\ 0.057,\ 0.051,\ 0.044$,  from up to down, respectively. We put  $\lambda=-1.5,\ \beta=0.14$.  With the decreasing of $Q$, the black holes make phases transition from 1-phase to 3-phase. }
\end{figure}

In Fig.~{\ref{Fig3f}} and Fig.~{\ref{Fig5f}}, we plot the $G(T)$ relations with running $\beta$ and $Q$, respectively. In Fig.~{\ref{Fig3f}}, the curves correspond to $\beta=0.2,\ 0.24,\ 0.28,\ 0.32,\ 0.335$, from outer to inner, respectively. We put  $M=1,\ Q=1$. Given $M,\ Q,\ \beta$, then $\lambda$ can be expressed as the function of $\lambda(M,Q,\beta,r_{+})$ by using the event horizon equation. Eventually, the Gibbs free energy $G$ becomes the parametric equations of $r_{+}$.  It shows with the increasing of $\beta$, the black holes make phases transition from 2-phase to 1-phase. It also tells us the Gibbs free energy of large black hole is decreasing with the increasing of Hawking temperature. On the other hand, the Gibbs free energy of small black hole first decreases and then increases with the increasing of Hawking temperature. As is known, the specific heat is $C_{Q,P}=-\partial^2G/\partial T^2$. Therefore, the thermodynamically stable and unstable phases have the concave downward and upward $G(T)$ curves, respectively. Then we conclude that the large black holes are thermodynamically stable while the small black holes are unstable.

In Fig.~\ref{Fig5f}, the curves correspond to  $Q=0.069,\ 0.063,\ 0.057,\ 0.051,\ 0.044$,  from up to down, respectively. We put  $\lambda=-1.5,\ \beta=0.14$. Given $\lambda,\ Q,\ \beta$, then $M$ can be expressed as the function of $M(\lambda,Q,\beta,r_{+})$ by using the event horizon equation. Eventually, the Gibbs free energy $G$ {becomes} the parametric equations of $r_{+}$. It shows that with the decreasing of $Q$, the black holes make phases transition from 1-phase to 3-phase. In this case, we let $Q$ be running. Then the Gibbs free energy of large, middle and small black holes all decreases with the increasing of Hawking temperature. But only large
and small black holes are thermodynamically  stable while the middle black holes are unstable.

%%%%%%%%%%%%%%%%%%%%%%%%%%%%%%%%%%%%
%%%%%%%%%%%%%%%%%%%%%%%%%%%%%%%%%%%%

\section{second solution}
In this section, we construct the second solution in Einstein-Maxwell-scalar theory. We remember that the dilaton black hole in de Sitter universe for arbitrary coupling constant $\alpha$ is \cite{gao:2004}
\begin{eqnarray}\label{eq:gao2}
U=\left(1-\frac{b}{r}\right)\left(1-\frac{a}{r}\right)^{\frac{1-\alpha^2}{1+\alpha^2}}-\frac{1}{3}\lambda f\;,
\end{eqnarray}
with
\begin{equation}
f=r^2\left(1-\frac{ a}{r}\right)^{\frac{2\alpha^2}{1+\alpha^2}}\;.
\end{equation}
Here $a,\ b$ are two constants which are determined by the black hole mass $M$, charge $Q$ and coupling constant $\alpha$. The corresponding coupling function $K$ and scalar potential $V$ in the action are
\begin{equation}
K=e^{2\alpha\phi}\;,
\end{equation}
and
\begin{eqnarray}\label{eq:2a}
V&=&\frac{2\lambda}{3\left(1+\alpha^2\right)^2}\left[\alpha^2\left(3\alpha^2-1\right)e^{2\phi/\alpha}+\left(3-\alpha^2\right)e^{-2\alpha\phi}\right.\nonumber\\&&\left.+8\alpha^2 e^{-\phi\alpha+\phi/\alpha}\right]\;.
\end{eqnarray}
Observing the $\lambda$ term in Eq.~(\ref{eq:gao2}), we find it is proportional to $f$. So we presume
\begin{equation}\label{eq:u2a}
U=\left(1-\frac{b}{r}\right)\left(1-\frac{a}{r}\right)^{\frac{1-\alpha^2}{1+\alpha^2}}+\frac{\beta Q^2}{f}-\frac{1}{3}\lambda f\;,
\end{equation}
\begin{equation}\label{eq:u2b}
f=r^2\left(1-\frac{a}{r}\right)^{\frac{2\alpha^2}{1+\alpha^2}}\;.
\end{equation}
We notice that when the coupling constant $\alpha=1$, the solution restores to the first solution in section III.
Substituting Eqs.~(\ref{eq:u2a}), (\ref{eq:u2b}) into the equations of motion (\ref{eom:1},\ref{eom:2},\ref{eom:3},\ref{eom:4}), we obtain
\begin{equation}\label{eq:2b}
\phi=-\frac{\alpha}{1+\alpha^2}\ln\left(1-\frac{a}{r}\right)\;,\ \ \ \ \  Q^2=\frac{ab}{\alpha^2+1}\;,
\end{equation}

\begin{equation}\label{eq:2c}
A_0=-\frac{Q}{r}-\frac{1}{1+\alpha^2}\frac{\beta Q}{r}-\frac{\alpha^2}{1+\alpha^2}\frac{\beta Q}{r-a}\;,
\end{equation}

\begin{equation}\label{eq:2d}
K=\frac{e^{2\alpha \phi}(\alpha^2+1)}{\alpha^2+\beta+1+\alpha^2 \beta e^{\frac{2\phi (\alpha^2+1)}{\alpha}}},
\end{equation}
with the scalar potential $V$ of the same expression in Eq.~(\ref{eq:2a}). When $\alpha=1$, they reduce to the counterparts in the first solution of section III.
So equations from (\ref{eq:2a}) to Eq.~(\ref{eq:2d}) constitute the second black hole solution in this study.

The scalar charge $D$ can be computed by
\begin{equation}
%D=\frac{1}{4\pi}\lim_{S^{\mu}\rightarrow i^{0}}\oint \nabla_{\mu}\phi \ dS^{\mu}=\frac{\gamma \alpha a}{\alpha^2 +1}\;.
D=-\frac{\alpha a}{\alpha^2 +1}\;.
\end{equation}

Here the integration is taken over the surface of spacetlike infinity, $i^{0}$. It tells us the scalar charge is determined by $m$, electric charge $Q$ {and} the coupling constant $\alpha$.

The electric charge of the black hole is the same as Eq.(\ref{charge}).
%\begin{equation}
%Q=\frac{1}{4\pi}\oint K \left({\phi}\right)\nabla_{\mu}A_{0}\ dS^{\mu}\;.
%\end{equation}
Here the integration is taken over a spacelike surface enclosing the origin. The electric charge is a conserved value and it does not depend on
the choice of the surface.

We find the ADM mass $M$ of the spacetime is
\begin{eqnarray}
%M&=&\frac{1}{16\pi}\lim_{S^{\mu}\rightarrow i^{0}}\oint g^{\alpha\beta}\left(g_{\alpha\mu,\beta}-g_{\alpha\beta,\mu}\right)dS^{\mu}\nonumber\\&=&\frac{1}{2}\left(b+\frac {-{\alpha}^{2}+1}{{\alpha}^{2}+1} a\right).
M=\frac{1}{2}\left(b+\frac {- {\alpha}^{2}+1}{{\alpha}^{2}+1} a\right).
\end{eqnarray}

The curvature singularity of this spacetime is at $r=a$. Similar to the first solution, there are at most three horizons in this spacetime and the corresponding causal structure is exactly the same as the Reissner-Nordstrom-de Sitter (or anti-de Sitter) spacetime. Therefore, we need not plot the Penrose diagram anymore. In the next section, we study the black hole thermodynamics.

\section{thermodynamics}
Following the procedure in section IV, we figure out the thermodynamic quantities for the second solution. We find the temperature, the electric potential and the thermal pressure are the same as Eq.(\ref{temperature}), Eq.(\ref{eq:p1}) and Eq.(\ref{eq:prs}), respectively.
%\begin{equation}\label{eq:t2}
%T=\frac{{\kappa}}{2\pi}=\frac{1}{4\pi}U^{'}\left(r_{+}\right)\;,
%T=\frac{1}{4\pi}U^{'}\left(r_{+}\right)\;,
%\end{equation}
The entropy and the thermal volume are
\begin{equation}
%S=\frac{A}{4}=\pi r_+^2\left(1-\frac{a}{r_+}\right)^{\frac{2\alpha^2}{1+\alpha^2}}\;,
S=\pi r_+^2\left(1-\frac{a}{r_+}\right)^{\frac{2\alpha^2}{1+\alpha^2}}\;,
\end{equation}
%the electric potential
%\begin{equation}\label{eq:p2}
%\Phi=-\int_{r_{+}}^{+\infty}A_0^{'}dr=A_0\left(r_{+}\right)\;,
%\end{equation}
%the thermal pressure
%\begin{equation}
%P=-\frac{\lambda}{8\pi}\;,
%\end{equation}
\begin{equation}\label{eq:v2}
\mathfrak{V}=\frac{4}{3}\,{\frac {\pi\, \left( -{\alpha}^{2}r_{+}+a-r_{+} \right) }{ \left( {\alpha
		}^{2}+1 \right) r_{+} \left( -r_{+}+a \right) }}
f^2\;.
\end{equation}
Then the Smarr formula Eq.(\ref{Smarr}) can be obtained.
In the following, we check whether these thermodynamic quantities satisfy the first law of thermodynamics.

From the equation of horizon, we obtain
\begin{equation}\label{eq:lamb}
\lambda=\frac{3}{f_+}\left[\left(1-\frac{b}{r_+}\right)\left(1-\frac{a}{r_+}\right)^{\frac{1-\alpha^2}{1+\alpha^2}}+\frac{\beta Q^2}{f_+}\right]\;,
\end{equation}
where $f_+=f\left(r=r_+\right)$. So the pressure $P$ can be written as
\begin{equation}
P=-\frac{3}{8\pi f_+}\left[\left(1-\frac{b}{r_+}\right)\left(1+\frac{a}{r_+}\right)^{\frac{1-\alpha^2}{1+\alpha^2}}+\frac{\beta Q^2}{f_+}\right]\;.
\end{equation}
We take the mass $M$, the pressure $P$, the entropy $S$ and the charge $Q$ as the function of $r_+,\ a$ and $b$. Then we derive the total differential of $P$, $S$, $Q$, $M$. By using the equations Eq.~(\ref{eq:t1}), Eq.~(\ref{eq:p1}) and Eq.~(\ref{eq:v2}), we arrive at the first law of thermodynamics Eq.(\ref{first law}).
\begin{figure}[htbp]
	\begin{center} \label{PVdiagram}
		\includegraphics[width=9cm]{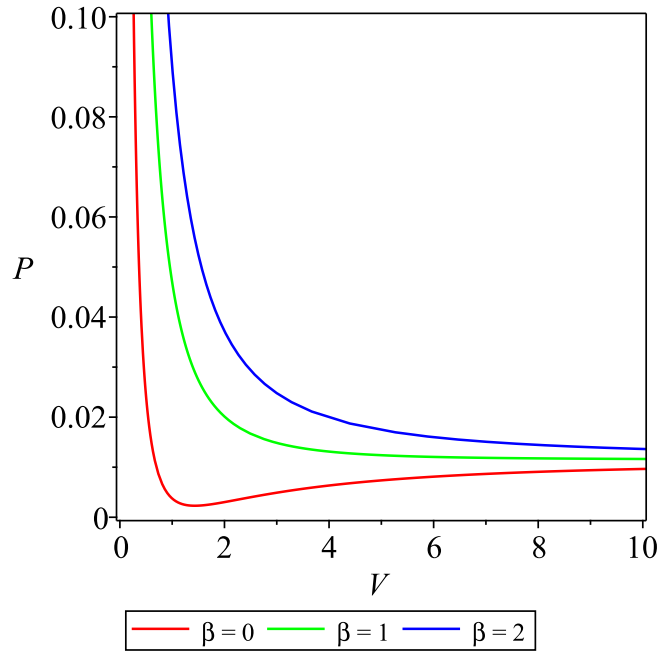}
		\caption{$P$ versus $V$ for $\alpha=0.1$, $Q=0.4$, $T=0.08$, and $\beta=0$ (red line), $\beta=1$ (green line), $\beta=2$ (blue line). }
	\end{center}
\end{figure}
\begin{figure}[htbp]
	\begin{center}\label{GTdiagram}
		\includegraphics[width=9cm]{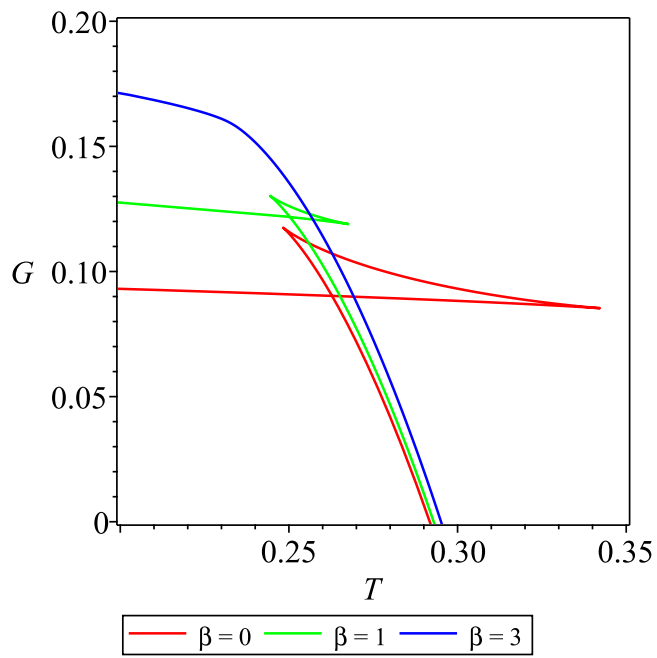}
		\caption{$G$ versus $T$ for $\alpha=0.1$, $Q=0.1$, $\lambda=-2.5$, and $\beta=0$ (red line), $\beta=1$ (green line), $\beta=3$ (blue line). }
	\end{center}
\end{figure}
To investigate the effect of $\beta$ on the black hole thermodynamical properties, we plot the $P-V$ diagram and $G-T$ diagram for fixed temperature and charge in Fig.~6 and Fig.~7. Fig.~6 shows that the isothermal curves are raised with the increasing of $\beta$. This means the temperature for Hawking-Page transition (critical temperature) decreases with the increasing of $\beta$. On the other hand, Fig.~7 shows that the temperature for Hawking-Page transition decreases with the increasing of $\beta$. Therefore, they are consistent with each other.

It has been shown by Sheykhi et al \cite{Sheykhi:2010} that for fixed values of the parameters in the dilaton-de Sitter black hole spacetime, there exists
a maximum value of the coupling constant $\alpha_{max}$ such that the black hole becomes thermally unstable. Compared to the dilaton-de Sitter solution, an extra coupling constant $\beta$ is in the presence in the second solution. Thus one may wonder if such a maximum value of the coupling constant $\alpha_{max}$ still remains. {It is the problem of local thermal stability for the system. For this point, following the method of Sheykhi et al, we investigate the variation of $\frac{\partial ^{2}M}{\partial S^{2}}$ with respect to $\alpha$ for different $\beta$. Fig.~(8) shows that for fixed $\beta$, the maximum value of the $\alpha_{max}$ remains. If $\alpha>\alpha_{max}$, $\frac{\partial ^{2}M}{\partial S^{2}}$ would become negative which means the system is locally and thermodynamically instable. But $\alpha_{max}$ increases with the increasing of $\beta$.

\begin{figure}[htbp]
	\begin{center}\label{capacity}
		\includegraphics[width=9cm]{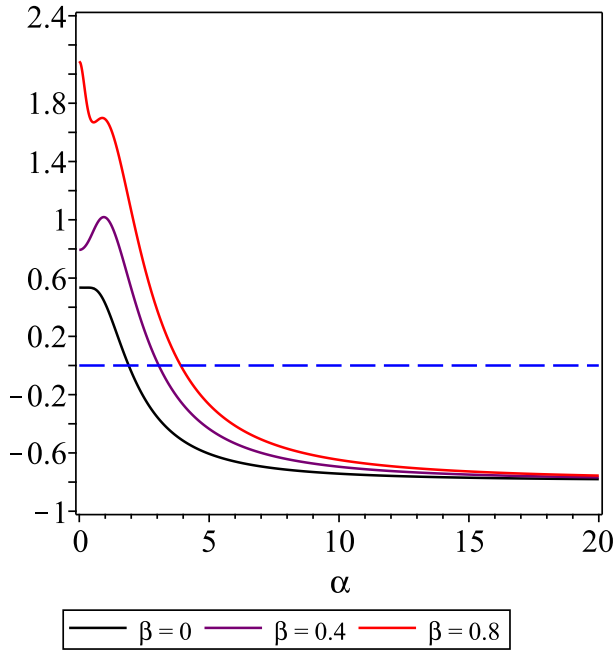}
		\caption{$\frac{\partial ^{2}M}{\partial S^{2}}$ versus $\alpha$ for $a=0.2$, $\lambda=-3$, $r_{+}=0.4$ and $\beta=0$ (black line), $\beta=0.4$ (purple line), $\beta=0.8$ (red line). }
	\end{center}
\end{figure}

\section{{Conclusion and discussion}}

In conclusion, we construct two exact and electrically charged black holes in the Einstein-Maxwell-scalar theory. The two solutions are all asymptotically de Sitter or anti-de Sitter. They are the extensions of dilaton black holes in de Sitter or anti-de Sitter universe. The first solution is for the case of $\alpha=1$ and the second solution is for arbitrary $\alpha$. In this sense, the second solution is the extension of the first one. An interesting fact is that the corresponding scalar potential $V(\phi)$ is the same as the dilaton counterpart. The only difference from the Einstein-Maxwell-dilaton theory is the presence of coupling function $K(\phi)$ in front of the Maxwell invariant. With the presence of $K(\phi)$, the well-known Reissner-Nordstrom-(anti) de Sitter solution is included in these solutions.

We calculate the ADM mass $M$, the electric charge $Q$ and the scalar charge $D$ of the black holes. The scalar charge $D$ is not independent, but dependent on $M$ and $Q$. This is the same as the dilaton black holes. Then we compute the Hawking temperature, the entropy, the electric potential and the thermal volume. Afterwards, both the Smarr formulae and the first law of thermodynamics are constructed. The evolution of Hawking temperature and Gibbs free energy shows that these black holes have two situations of phase structure, the scenario of two phases and three phases. In the scenario of two phases, we let the coupling constants $\beta$ and $\alpha$ be running. In this case, the Gibbs free energy of large black hole is decreasing with the increasing of Hawking temperature. On the other hand, the Gibbs free energy of small black holes first decreases and then increases with the increasing of Hawking temperature. This shows the large black hole is thermodynamically stable and the small black hole is unstable. In the scenario of three phases, we let $Q$ be running. In this case, the Gibbs free energy of large, middle and small black holes all decreases with the increasing of Hawking temperature. But only the large and small black holes are thermodynamically stable while the middle black holes are unstable. Sheykhi et al \cite{Sheykhi:2010} have shown that, for fixed values of the parameters in the dilaton-de Sitter black hole spacetime, there exists
a maximum value of the coupling constant $\alpha_{max}$ such that the black hole becomes thermally unstable. Compared to the dilaton-de Sitter solution, an extra coupling constant $\beta$ is in presence in our new solution. We find that such a maximum value of the coupling constant $\alpha_{max}$ still remains in the new solution. But $\alpha_{max}$ increases with the increasing of $\beta$.

{Finally, since these black hole solutions are asymptotically anti-de Sitter, it is worth studying the corresponding ADS/CFT correspondence. On the other hand, the extension of the static black holes to the rotating and higher dimensional cases {can be a topic} worthy of study.}
\section*{ACKNOWLEDGMENTS}

We thank the anonymous referees for constructive comments. This work is partially supported by the Strategic Priority Research Program ``Multi-wavelength Gravitational Wave Universe'' of the CAS, Grant No. XDB23040100 and the NSFC under grants 11633004, 11773031.

%\section*{References}

\newcommand\ARNPS[3]{~Ann. Rev. Nucl. Part. Sci.{\bf ~#1}, #2~ (#3)}
\newcommand\AL[3]{~Astron. Lett.{\bf ~#1}, #2~ (#3)}
\newcommand\AP[3]{~Astropart. Phys.{\bf ~#1}, #2~ (#3)}
\newcommand\AJ[3]{~Astron. J.{\bf ~#1}, #2~(#3)}
\newcommand\GC[3]{~Grav. Cosmol.{\bf ~#1}, #2~(#3)}
\newcommand\APJ[3]{~Astrophys. J.{\bf ~#1}, #2~ (#3)}
\newcommand\APJL[3]{~Astrophys. J. Lett. {\bf ~#1}, L#2~(#3)}
\newcommand\APJS[3]{~Astrophys. J. Suppl. Ser.{\bf ~#1}, #2~(#3)}
\newcommand\JHEP[3]{~JHEP.{\bf ~#1}, #2~(#3)}
\newcommand\JMP[3]{~J. Math. Phys. {\bf ~#1}, #2~(#3)}
\newcommand\JCAP[3]{~JCAP {\bf ~#1}, #2~ (#3)}
\newcommand\LRR[3]{~Living Rev. Relativity. {\bf ~#1}, #2~ (#3)}
\newcommand\MNRAS[3]{~Mon. Not. R. Astron. Soc.{\bf ~#1}, #2~(#3)}
\newcommand\MNRASL[3]{~Mon. Not. R. Astron. Soc.{\bf ~#1}, L#2~(#3)}
\newcommand\NPB[3]{~Nucl. Phys. B{\bf ~#1}, #2~(#3)}
\newcommand\CMP[3]{~Comm. Math. Phys.{\bf ~#1}, #2~(#3)}
\newcommand\CQG[3]{~Class. Quantum Grav.{\bf ~#1}, #2~(#3)}
\newcommand\PLB[3]{~Phys. Lett. B{\bf ~#1}, #2~(#3)}
\newcommand\PRL[3]{~Phys. Rev. Lett.{\bf ~#1}, #2~(#3)}
\newcommand\PR[3]{~Phys. Rep.{\bf ~#1}, #2~(#3)}
\newcommand\PRd[3]{~Phys. Rev.{\bf ~#1}, #2~(#3)}
\newcommand\PRD[3]{~Phys. Rev. D{\bf ~#1}, #2~(#3)}
\newcommand\RMP[3]{~Rev. Mod. Phys.{\bf ~#1}, #2~(#3)}
\newcommand\SJNP[3]{~Sov. J. Nucl. Phys.{\bf ~#1}, #2~(#3)}
\newcommand\ZPC[3]{~Z. Phys. C{\bf ~#1}, #2~(#3)}
\newcommand\IJGMP[3]{~Int. J. Geom. Meth. Mod. Phys.{\bf ~#1}, #2~(#3)}
\newcommand\IJMPD[3]{~Int. J. Mod. Phys. D{\bf ~#1}, #2~(#3)}
\newcommand\IJMPA[3]{~Int. J. Mod. Phys. A{\bf ~#1}, #2~(#3)}
\newcommand\GRG[3]{~Gen. Rel. Grav.{\bf ~#1}, #2~(#3)}
\newcommand\EPJC[3]{~Eur. Phys. J. C{\bf ~#1}, #2~(#3)}
\newcommand\PRSLA[3]{~Proc. Roy. Soc. Lond. A {\bf ~#1}, #2~(#3)}
\newcommand\AHEP[3]{~Adv. High Energy Phys.{\bf ~#1}, #2~(#3)}
\newcommand\Pramana[3]{~Pramana.{\bf ~#1}, #2~(#3)}
\newcommand\PTP[3]{~Prog. Theor. Phys{\bf ~#1}, #2~(#3)}
\newcommand\APPS[3]{~Acta Phys. Polon. Supp.{\bf ~#1}, #2~(#3)}
\newcommand\ANP[3]{~Annals Phys.{\bf ~#1}, #2~(#3)}
\newcommand\RPP[3]{~Rept. Prog. Phys. {\bf ~#1}, #2~(#3)}

\end{document}